%% file: acl2023.tex
\pdfoutput=1

\documentclass[11pt,dvipsnames]{article}

\usepackage[]{ACL2023}

\usepackage{times}
\usepackage{latexsym}

\usepackage[T1]{fontenc}

\usepackage[utf8]{inputenc}

\usepackage{microtype}

\usepackage{inconsolata}

\usepackage{multirow,hyperref}
\usepackage{graphicx}
\usepackage{diagbox}
\usepackage{booktabs}
\usepackage{xspace}
\usepackage[ruled, vlined, linesnumbered]{algorithm2e}
\usepackage{bbding}
\usepackage{threeparttable}
\usepackage{amsmath}
\usepackage{adjustbox}
\usepackage{listings}
\usepackage[T1]{fontenc}
\usepackage{semantic}
\usepackage{newfloat}
\usepackage{framed}
\usepackage{threeparttable}

\usepackage{makecell}

\usepackage{hyperref}
\hypersetup{
colorlinks=true,
linkcolor=blue,
filecolor=megenta
}

\newcounter{mydoublefinding}
\counterwithin*{mydoublefinding}{subsection}
\newcounter{myfindings}
\counterwithin*{myfindings}{section}
\usepackage{pifont}

\newcommand{\CodeIn}[1]{{\small\texttt{#1}}}

\DeclareFloatingEnvironment[fileext=frm,placement={!ht},name=Rules]{infrule}

\usepackage{algpseudocode}
\usepackage[normalem]{ulem}
\useunder{\uline}{\ul}{}

\newcommand\detector{VulLibGen}

\title{VulLibGen: Generating Names of Vulnerability-Affected Packages via a Large Language Model}

\author{
  Tianyu Chen\textsuperscript{1}, Lin Li\textsuperscript{2}, Liuchuan Zhu\textsuperscript{2}, Zongyang Li\textsuperscript{1}, Xueqing Liu\textsuperscript{3}\\
  \textbf{Guangtai Liang\textsuperscript{2}, Qianxiang Wang\textsuperscript{2}, Tao Xie\textsuperscript{1}$^{\ast}$} \\
  Key Lab of HCST (PKU), MOE; SCS, Peking University\textsuperscript{1}\\
  \texttt{\{tychen811,taoxie\}@pku.edu.cn, lizongyang@stu.pku.edu.cn} \\
  Huawei Cloud Computing, Technologies Co., Ltd.\textsuperscript{2} \\
  \texttt{\{lilin88,zhuliuchuan1,liangguangtai,wangqianxiang\}@huawei.com} \\
  Stevens Institute of Technology\textsuperscript{3}, \texttt{xliu127@stevens.edu}\\
\thanks{*Corresponding author}
}

\begin{document}
\maketitle
\begin{abstract}

Security practitioners maintain vulnerability reports (e.g., GitHub Advisory) to help developers mitigate security risks. An important task for these databases is automatically extracting structured information mentioned in the report, e.g., the affected software packages, to accelerate the defense of the vulnerability ecosystem. 

However, it is challenging for existing work on affected package identification to achieve a high accuracy. One reason is that all existing work focuses on relatively smaller models, thus they cannot harness the knowledge and semantic capabilities of large language models. 

To address this limitation, we propose \detector{}, the first method to use LLM for affected package identification. 
In contrast to existing work, \detector{} proposes the novel idea to directly generate the affected package. To improve the accuracy, \detector{} employs supervised fine-tuning (SFT), retrieval augmented generation (RAG) and a local search algorithm. The local search algorithm is a novel post-processing algorithm we introduce for reducing the hallucination of the generated packages.  
Our evaluation results show that \detector{} has an average accuracy of 0.806 for identifying vulnerable packages in the four most popular ecosystems in GitHub Advisory (Java, JS, Python, Go) while the best average accuracy in previous work is 0.721. 
Additionally, \detector{} has high value to security practice:  
we submitted 60 <vulnerability, affected package> pairs to GitHub Advisory (covers four ecosystems). 34 of them have been accepted and merged and 20 are pending approval. Our code and dataset can be found in the attachments.

 \end{abstract}

\input{introduction}
\input{background}

\input{approach}
\input{evaluation}
\input{related}

\input{conclusion}

\input{limitations}

\bibliographystyle{acl_natbib}
\bibliography{sample-base}
\input{appendix}

\end{document}

%% file: introduction.tex
\begin{figure}[h]
\centering
\includegraphics[width=1\linewidth]{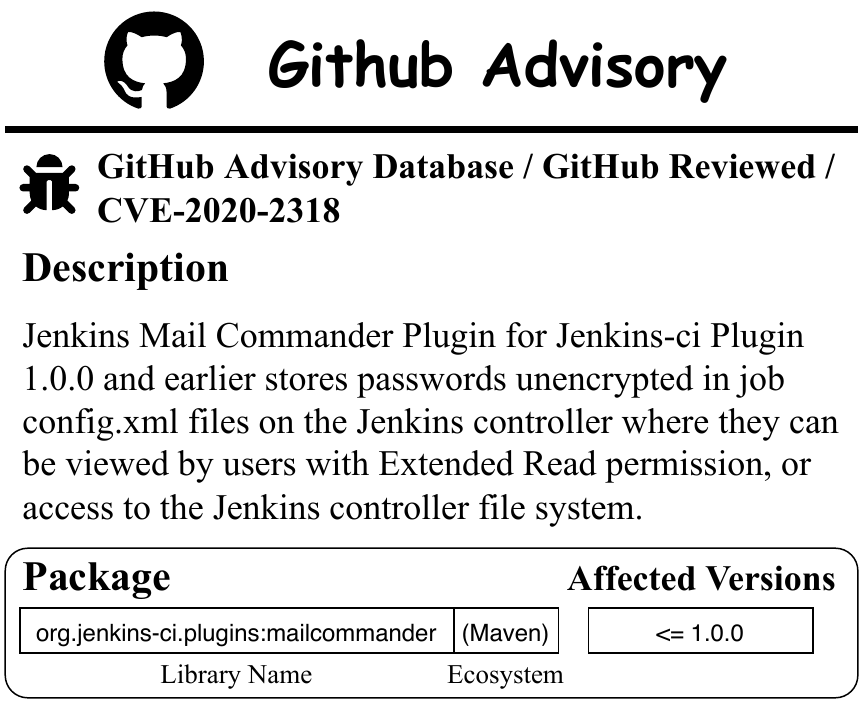}
\caption{GitHub Advisory Report for \href{https://github.com/advisories/GHSA-485q-v457-3p58}{CVE-2020-2318}}
\label{fig: case}
\end{figure}

\section{Introduction}
\label{sec:intro}
One important task in security is to automatically extract the package names and versions affected by each security vulnerability. 
A recent study~\cite{wang2020empirical} shows that 84\% third-party packages contain security vulnerabilities and 60\% of them are high-risk ones.
To mitigate the security risks, security practitioners maintain databases (e.g., NVD and GitHub Advisory) for each unique vulnerability (i.e., Common Weakness Enumeration or CVE). These databases contain reports with the vulnerability description, affected packages, and affected versions. For example, Figure~\ref{fig: case} shows the report of \href{https://github.com/advisories/GHSA-485q-v457-3p58}{CVE-2020-2318}. By linking the CVEs to the affected packages, developers can be aware of the vulnerable packages in their code and quickly apply patches/fixes; the affected packages also help with security knowledge management and task prioritization. Although developers can manually enter the affected packages when submitting CVEs, the entered information is often missing or incorrect~\cite{lightxml,viem}. While database maintainers also conduct manual reviews~\cite{githubReview}, automatic identification can accelerate the vulnerability lifecycle and reduce the manual costs, thus it is a critical task. 


Several existing works have studied automatic affected package identification~\cite{fastxml, viem, lightxml, chronos, vullibminer}, however, it is challenging for them to achieve high accuracy because they fail to leverage large language models. Existing works typically rank/retrieve the package name from a list of pre-defined packages by computing the similarity between the vulnerability description and the package description. Since the time complexity of retrieval is linear to the size of the list (e.g., 435k Java packages), the cost of each model inference has to be kept quite low and only smaller models have been used, e.g., BERT and linear regression~\cite{fastxml,chronos,lightxml,vullibminer}. 

To leverage the extensive knowledge and semantic capabilities of LLMs, we propose a different strategy: \emph{generate rather than retrieve} the affected package. Our framework, \detector{}, is the first framework to directly generate the mentioned package name given the vulnerability description. Since generation requires the LLM inference to be invoked only once, our approach can easily scale to larger language models such as Llama-13B and Vicuna-13B on a single GPU server. 

Upon observing errors in the raw generation result of LLMs, we propose the following techniques for improving \detector{}'s accuracy. 
First, we conduct supervised fine-tuning (SFT), in-context learning, and retrieval-augmented generation (RAG) to enhance the domain knowledge of a LLM.
Second, we propose a novel local search technique to post-process the raw output for reducing hallucination, i.e., ensuring that the generated package name actually exists. Our local search algorithm is inspired by an empirical study of the incorrect raw outputs of ChatGPT. The study shows that most errors are partially correct, therefore we can create rules to match the incorrect part to the closest existing one given the correct part. Since the sub-level package information (e.g., \CodeIn{mailcommander} in Figure~\ref{fig: case}) is often more directly mentioned than the top-level package information (e.g., \CodeIn{org.jenkins-ci.plugins}), our local search algorithm first matches the suffix before the prefix.

We evaluate \detector{} on four vulnerability ecosystems: Java, JS, Python and Go. Our evaluation attains three main findings. First, we observe that the accuracy of \detector{} (0.806) significantly outperforms existing ranking approaches using smaller models~\cite{fastxml,lightxml,chronos,vullibminer} (0.721) and the computational time costs are comparable. 
Second, our ablation studies show that SFT, RAG, and local search all help improve the accuracy of \detector{} and SFT contributes to the most improvement. In particular, the fine-tuned open-source Vicuna-13B outperforms the unfine-tuned commercial ChatGPT and GPT-4 models. Our local search algorithm can significantly reduce the hallucination in the original LLM output, and it is especially helpful for longer package names such as Java and Go. Third, \detector{} provides high value to security practice: at the time of the writing, we have submitted 60 pairs of <vulnerability, affected package> to GitHub Advisory (25 Java, 14 JS, 11 Python, 10 Go). 34 of them are accepted and merged and 20 are pending approval. 
Additionally, the source code and dataset of \detector{} is available on our anonymous website~\footnote{https://github.com/anonymous4ACL24/submission1129}.

%% file: background.tex
\begin{table*}[t]
    \centering
    \caption{An Empirical Study on ChatGPT's Incorrect Response in Maven (Java ecosystem)}
    \label{tab: fault case study}
\scriptsize
\begin{threeparttable}
\begin{tabular}{p{2cm}p{1.8cm}p{4.3cm}p{4.3cm}}
\toprule
\textbf{Error Reason}                                            & Example (w/ link)                             & ChatGPT's Output                                                   & Ground Truth (Affected Packages)                                                         \\
\midrule
\multirow{2}{*}{\makecell[l]{\textbf{Type 1: Incorrect}\\ \textbf{but exist (23\% of}\\ \textbf{ all errors)}}} & \href{https://github.com/advisories/GHSA-9qhq-j4xm-cw48}{CVE-2015-3158}                     & org.\textcolor{Blue}{picketlink:picketlink}                                    & org.\textcolor{Blue}{picketlink:picketlink-tomcat}-common                                     \\
&&  \multicolumn{2}{l}{\textit{Description:} ``The invokeNextValve function in \textit{identity/federation/bindings/\textcolor{Blue}{tomcat}/idp/AbstractIDPValve.java}}                             \\
&& \multicolumn{2}{l}{in \textcolor{Blue}{PicketLink} before 2.7.1.Final does not properly check role based authorization, which allows remote} \\
&& \multicolumn{2}{l}{authenticated users to gain access to restricted application resources via a (1) direct request $\dots$'' } \\
\midrule
\multirow{8}{*}{\makecell[l]{\textbf{Type 2: Non-Exist,}\\ \textbf{Partially correct}\\ \textbf{(58\% of all errors)}}}                                          & \href{https://github.com/advisories/GHSA-wv88-pf73-x22p}{CVE-2011-2730}                     & org.\textcolor{Blue}{springframework}:\textcolor{Red}{spring-framework}                                    & org.\textcolor{Blue}{springframework}:\textcolor{Blue}{spring-core}                                      \\
&&  \multicolumn{2}{l}{\textit{Description:} ``VMware \textcolor{Blue}{SpringSource Spring Framework} before 2.5.6.SEC03, 2.5.7.SR023, and 3.x before  }                                                                                                                                                                                \\
&& \multicolumn{2}{l}{ 3.0.6, when a container supports Expression Language (EL), evaluates EL expressions in tags twice which } \\
& & \multicolumn{2}{l}{allows remote attackers to obtain sensitive information. $\dots$''}                                                                                                                                                                                \\
\cmidrule(lr){3-4}
  & \href{https://github.com/advisories/GHSA-264w-xrr7-6qqg}{CVE-2020-2167}                       & \textcolor{Red}{org.jenkins-ci.plugins}:\textcolor{Blue}{openshift-pipeline}                    & \textcolor{Blue}{com.openshift.jenkins}:\textcolor{Blue}{openshift-pipeline}                       \\
&&\multicolumn{2}{l}{\textit{Description:} ``\textcolor{Blue}{OpenShift Pipeline Plugin} 1.0.56 and earlier does not configure its YAML parser to prevent} \\
&&\multicolumn{2}{l}{
the instantiation of arbitrary types. This results in a remote code execution (RCE) vulnerability exploitable}                                                                                                                                                                      \\
&& \multicolumn{2}{l}{  by users able to provide YAML input files to OpenShift Pipeline Plugin’s build step. $\dots$''}                                                                                                                                                                      \\
\midrule
\multirow{4}{*}{\makecell[l]{\textbf{Type 3: Non-Exist,}\\ \textbf{Completely incorrect}\\ \textbf{(19\% of all errors)}}}&                      \href{https://github.com/advisories/GHSA-jpj4-5xwp-cv23}{CVE-2020-11974}                      & \textcolor{Red}{mysql:mysql-connector-java}                                   & \textcolor{Blue}{org.apache.dolphinscheduler:dolphinscheduler}                   \\
&& \multicolumn{2}{l}{\textit{Description:} ``In \textcolor{Blue}{DolphinScheduler} 1.2.0 and 1.2.1, with \textcolor{Red}{mysql connector} a remote code execution vulnerab}\\
&& \multicolumn{2}{l}{-ility
exists when choosing mysql as database.''}                          \\
\cmidrule(lr){3-4}
&          \href{https://github.com/advisories/GHSA-fxp8-7h5w-h235}{CVE-2019-13234}                      & \textcolor{Red}{N/A}                                                                  & \textcolor{Blue}{org.opencms:opencms-core}                                       \\
&&\multicolumn{2}{l}{\textit{Description:} ``In the Alkacon \textcolor{Blue}{OpenCms} Apollo Template 10.5.4 and 10.5.5 there is XSS in the search engine.''}  \\
\bottomrule
\end{tabular}
\end{threeparttable}
\vspace{-0.3cm}
\end{table*}

\section{Existing Work on Vulnerable Package Identification\label{sec: scale}}

This section summarizes existing work on affected package identification and analyze the scalability challenge. 

\noindent \textbf{Formal Definition of Affected Package Identification}. Given a security vulnerability (CVE) submitted to a software ecosystem (e.g., GitHub Advisory), the goal of affected package identification is to link the description $q$ of the CVE to an existing software package name $p$ (e.g., a Maven or PyPi package) that is affected by the CVE. An example of the linked package can be found in Figure~\ref{fig: case}, where the description mentions the affected package \CodeIn{org.jenkins-ci.plugins:mailcommander}. 


\noindent \textbf{Smaller Models Have Lower Accuracies}. 
Existing approaches on vulnerable package identification~\cite{fastxml, viem, lightxml, chronos} all suffer from lower accuracies~\cite{chronos,vullibminer}. Given the vulnerability $q$, existing works rank all packages $p$ of the ecosystem by computing the similarity score between the descriptions of $q$ and $p$. Due to the large number of packages (e.g., Maven has 435k packages), existing work cannot afford using large language model to compute the score. All of them thus rely on smaller models, e.g., logistic regression~\cite{zestxml,chronos} and BERT~\cite{lightxml,vullibminer}. Despite various methods introduced for improving the accuracy~\cite{viem, anwar2021cleaning}, the accuracies remain low~\cite{vullibminer}.

\noindent \textbf{Existing Work's Efforts on Scaling to Larger Models}. To improve the accuracy, existing work leverages re-ranking with the BERT model~\cite{vullibminer}. More specifically, they first use TF-IDF to rank all packages in the ecosystem (435k in Java and 506k in Python), then re-rank the top-512 packages using BERT. The re-ranking approach achieves a reasonable accuracy with a saved inference cost, but there remains a large room for improving the accuracy~\cite{vullibminer}.

\section{Two Challenges with LLM Generation~\label{sec: empirical_study}}


In contrast to existing work, we propose the first work, \detector{}, that leverages LLMs for affected package identification. Due to the scalability challenge of the retrieval approach, our approach directly \emph{generates rather than retrieves} the affected package. \detector{} thus only need to invoke the LLM inference once for each vulnerability $q$.  
Nevertheless, there exist two challenges with the generative approach. 

\noindent \textbf{Challenge 1: Lack of Domain Knowledge}. The first challenge is that there may exist a knowledge gap for the LLM to generate the correct package. This is because the description may not contain the full information about the affected package name. For example, \href{https://github.com/advisories/GHSA-264w-xrr7-6qqg}{CVE-2020-2167} in Table~\ref{tab: fault case study} is about the Java package \CodeIn{com.openshift.jenkins.openshift-pipeline}, but the the description does not mention the word "\emph{Jenkins}". To predict the correct package name, the LLM has to rely on domain knowledge to complete this information. Existing work have used various methods to bridge the knowledge gap of LLMs, e.g., supervised fine-tuning~\cite{prottasha2022transfer, church2021emerging} and retrieval augmented generation~\cite{lewis2020retrieval,
mao2020generation, liu2020retrieval, cai2022recent}.

\noindent \textbf{Challenge 2: Generating Non-Existing Package Names}. Following a previous study on Reddit\footnote{\url{https://www.reddit.com/r/ChatGPT/comments/zneqyp/chatgpt\_hallucinates\_a\_software\_library\_that/}}, the second challenge is that the LLM may generate library names that do not exist in the ecosystem. Existing work has adopted post-processing to reduce the non-existing package issue in code translation and program repair~\cite{jin2023inferfix, roziere2021leveraging}. Following existing work, we can potentially leverage post-processing by matching the generated package with the closest existing package.  

To understand whether post-processing is promising for solving Challenge 2 and to study how to design the post-processing algorithm, we conduct an empirical study on ChatGPT's incorrect response, the study result can be seen in Table~\ref{tab: fault case study}. The study uses 2,789 Java vulnerability descriptions collected in a recent work~\cite{vullibminer}. We divide all ChatGPT responses into four types: 1. the package is incorrect but it exists (13\%, 23\% of errors); 2. the package does not exist and is partially correct (34\%, 58\% of errors); 3. the package is completely incorrect (11\%, 19\% of errors). 4. the package is correct (42\% of all cases); 

\begin{figure*}[t]
\centering
\includegraphics[width=1\linewidth]{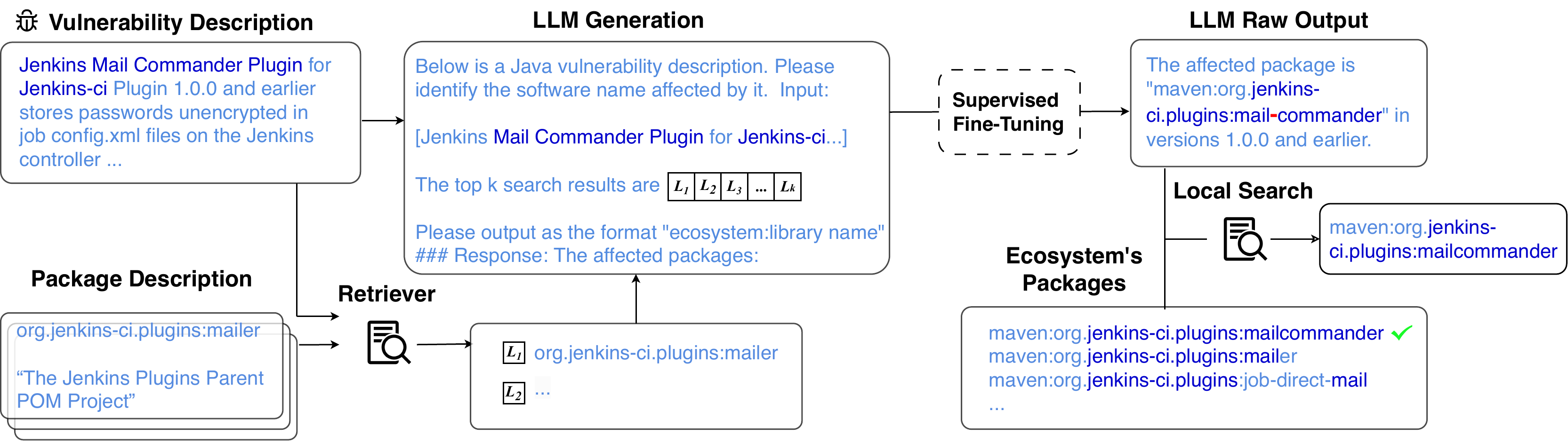}
\caption{The \detector{} Framework}
\label{fig: framework}
\vspace{-0.3cm}
\end{figure*}

From the study result, we draw the conclusion that post-processing by matching is a promising approach to solve Challenge 2. This is because the majority errors are Type 2 errors, while post-processing is the most effective in helping with Type 2 errors. For example, for \href{https://github.com/advisories/GHSA-264w-xrr7-6qqg}{CVE-2020-2167}, ChatGPT generates \textcolor{Red}{org.jenkins-ci.plugins}:\textcolor{Blue}{openshift-pipeline}. While the \textcolor{Blue}{suffix} is correct, the \textcolor{Red}{prefix} and \textcolor{Blue}{suffix} never co-occur in any existing package name. We can fix this case by matching the prefix to the closest co-occured one. 
By applying a naive edit-distance matching on the ChatGPT output, the accuracy is improved from 42\% to 51\%.

%% file: approach.tex
\section{\detector{} Framework}\label{sec:approach}

To address the two challenges in LLM generation, we employ the following techniques: first, we use supervised fine-tuning and in-context learning to enhance the domain knowledge in LLM; second, we further employ the retrieval-augmented framework (RAG) to enhance the knowledge when SFT is not easy; third, we design a local search technique which alleviates the non-existing package name problem. The \detector{} framework can be found in Figure~\ref{fig: framework} \footnote{Our prompt in Figure~\ref{fig: framework} is: "\emph{Below is a [Programming Language] vulnerability description. Please identify the software name affected by it. Input: [DESCRIPTION]. The top k search results are: [$L_1$][$L_2$]$\cdots$[$L_k$]. Please output the package name in the format "ecosystem:library name".
\#\#\# Response: The affected packages:}". ~\label{fn: input}}.

\subsection{Supervised Fine-Tuning/In-Context Learning}

To solve the first challenge (Section~\ref{sec: empirical_study}), we incorporate supervised fine-tuning~\cite{prottasha2022transfer, church2021emerging} and in-context learning~\cite{dong2022survey, olsson2022context} in \detector{}. For SFT, we use the full training data (Table~\ref{tab: dataset info}); for ICL, we randomly sample 3 examples from the training data for each evaluation vulnerability. For both SFT and ICL, the input and output of the LLM follow the following format: Input: the same prompt as Figure~\ref{fig: framework}~\footref{fn: input}, Output: "\emph{The affected package is [package name]}". The hyper-parameters used for ICL and SFT are listed in Table~\ref{tab: hyper parameter} of Appendix.


\subsection{Retrieval-Augmented Generation (RAG)}

To further enhance the LLM's domain knowledge especially when SFT is not easy (e.g., ChatGPT and GPT4), we employ retrieval-augmented generation (RAG) in \detector{}.

\noindent \textbf{Retriever setting}. Given the description of a vulnerability, our retriever ranks existing package names in an ecosystem (Table~\ref{tab: dataset info}) based on the similarity score between the vulnerability description and the package description. The descriptions of Java, JavaScript, Python, and Go packages are obtained from Maven~\footnote{\url{https://mvnrepository.com}}, NPM~\footnote{\url{https://www.npmjs.com}}, Pypi~\footnote{\url{https://pypi.org}}, and Go~\footnote{\url{https://pkg.go.dev}} documentations.
For example, the description of the package \CodeIn{org.jenkins-ci.plugins:mailcommander} is \textit{``This plug-in provides function that read a mail subject as a CLI Command.''}.
Our retriever follows \cite{vullibminer}'s re-ranking strategy, i.e., first rank all packages (e.g., 435k in Java) using TF-IDF, then re-rank the top 512 packages using a BERT-base model fine-tuned on the same training data in Table~\ref{tab: dataset info}.


\subsection{Local Search}

To solve the second challenge (Section~\ref{sec: empirical_study}), we incorporate post-processing in \detector{}.
Based on the empirical study results in Section~\ref{sec: empirical_study}, we design a local search technique to match the generation output with the closest package name from an existing package list (Algorithm~\ref{alg: local search} in Appendix~\ref{sec: local search algorithm}). 

Algorithm~\ref{alg: local search} employs the edit distance as the metric and respects the structure of the package name. Formally, a package name can be divided into two parts: its prefix and suffix (separated by a special character, e.g., `:' in Java).
The prefix (e.g., the artifact ID of Java packages) specifies the maintainer/group of this package, and the suffix (e.g., the group ID of Java packages) specifies the functionalities of this package.
Specifically, Java, Go, and part of JS packages can be explicitly divided while Python and the rest of JS packages only specify their functionalities in their names.
We denote the prefix of a package name as empty if it can not be divided.

Algorithm~\ref{alg: local search} first compares the generated suffix with all existing suffix names and matches the suffix to the closest one. After fixing the suffix, we can then obtain the list of prefixes that co-occur at least once with this suffix. We match the generated prefix with the closest prefix in this list.
The reason that we opt to match the suffix first is twofold. First, our study shows that the vulnerability description more frequently mentions the suffix than the prefix: among all 2,789 vulnerabilities investigated in Section~\ref{sec: empirical_study}, their description mentions 12.4\% of the tokens in the prefixes of the affected packages and 66.0\% of the tokens in their suffixes of the affected packages;
second, our study also shows that each suffix co-occurs with fewer unique prefixes than conversely.
In all 435k Java packages, each prefix has 5.86 co-occurred suffixes while each suffix has only 1.13 co-occurred prefixes on average.
As a result, it is easier to identify the prefix by first matching the suffix, and then matching the suffix with the co-occurred prefix list. 

%% file: evaluation.tex
\section{Evaluation}
\label{sec:evalution}

\subsection{Evaluation Setup}

\noindent \textbf{Dataset}. Among existing work on affected package identification~\cite{fastxml,lightxml,chronos,vullibminer}, two datasets are frequently used: VeraCode~\cite{fastxml} and VulLib~\cite{vullibminer}. In this work, we choose to use VulLib instead of VeraCode because it is of better quality. The VulLib dataset contains 2,789 Java vulnerabilities collected from GitHub Advisory. Each package name in VulLib is manually verified by security experts from GitHub Advisory~\cite{githubAD}. In contrast, VeraCode is not verified thus it is prone to errors~\footnote{VeraCode does not use explicit package names ecosystems. For example, the affected package of CVE-2014-2059 is ``org.jenkins-ci.main:jenkins-core'' while VeraCode labels it as three packages: ``jenkins'', ``openshift-origin-cartridge-jenkins'', and ``jenkins-plugin-openshift''}.
Since VulLib only focuses on Java, we further extend it to JS, Python and Go by collecting the data from GitHub Advisory following a similar workflow as VulLib.

The statistics of our dataset are listed in Table~\ref{tab: dataset info}. In total, our dataset includes 2,789 Java, 3,193 JS, 2,237 Python, and 1,351 Go vulnerabilities, respectively. To the best of our knowledge, this is the first dataset for identifying vulnerable packages with various programming languages. For each PL, we split the train/validation/test data with the 3:1:1 ratio. The split is in chronological order to simulate a more realistic scenario and to prevent lookahead bias~\cite{lookahead_bias}.



\begin{table}[t]
\centering
\caption{The Statistics of the GitHub Advisory Dataset}
\label{tab: dataset info}
\begin{threeparttable}
\small
\begin{tabular}{lrrrr}
\toprule
           & Java & JS   & Python & Go   \\
\midrule
\multicolumn{5}{l}{\textit{\#Vulnerabilities:}}     \\
Training   & 1,668 & 1,915 & 1,342   & 810   \\
Validation & 556   & 639   & 447     & 270   \\
Testing    & 565   & 639   & 448     & 271   \\
Total      & 2,789 & 3,193 & 2,237   & 1,351 \\
\midrule
\multicolumn{5}{l}{\textit{\#Unique packages in the dataset:}}    \\
      & 2,095 & 2,335 & 710     & 601   \\ 
\multicolumn{5}{l}{\textit{\#Total packages in their ecosystems:}}    \\
      & 435k & 2,551k & 507k     & 12k   \\ 
\midrule
\multicolumn{5}{l}{\textit{\#Avg. tokens of packages:}}    \\
& 13.44 & 4.56  & 3.96    & 8.24 \\
\bottomrule
\end{tabular}
\end{threeparttable}
\end{table}

\noindent \textbf{Comparative Methods}. To evaluate the effectiveness of \detector{}, we contrast it with four existing ranking approaches, FastXML~\cite{fastxml}, LightXML~\cite{lightxml}, Chronos~\cite{chronos}, and VulLibMiner~\cite{vullibminer} for comparison.
Recent studies~\cite{chronos, vullibminer} show that they outperform other approaches, such as Bonsai~\cite{khandagale2020bonsai} and ExtremeText~\cite{wydmuch2018no}.

\noindent \textbf{Models in \detector{}}. The models we evaluate for the \detector{} framework include both commercial LLMs, e.g., ChatGPT (gpt-3.5-turbo) and GPT4 (gpt-4-1106-preview), and open-source LLMs, e.g., LLaMa~\cite{llama} and Vicuna~\cite{vicuna}.

We assess open-source LLMs in two scenarios: few-shot in-context learning using 3 examples\
randomly sampled from the training data and supervised fine-tuning using the full training data.
For the open-source LLMs, we use ICL/SFT + RAG + local search, whereas for commercial LLMs, we use RAG + local search only. 




\noindent \textbf{Evaluation Environments}
Our evaluations are conducted on the system of Ubuntu 18.04. 
We use one Intel(R) Xeon(R) Gold 6248R@3.00GHz CPU, which contains 64 cores and 512GB memory.
We use 8 Tesla A100 PCIe GPUs with 40GB memory for model training and inference. 
In total, our experiments constitute 200 GPU days (32 groups in RQ1 + 68 groups in RQ2, and each group costs 0.25 GPU days across 8 GPUs).

\noindent \textbf{Metrics}
Following previous work~\cite{vullibminer, fastxml}, we use three metrics for evaluating \detector{} and baselines: Accuracy@1 (i.e., Prec@1), Recall@1, and F1@1. 


\subsection{Evaluation of \detector{}~\label{sec: trade off}}

In this subsection, we evaluate the effectiveness of \detector{}. We seek to answer the following research question: How does \detector{} compare to existing work on identifying vulnerable packages? 

\begin{table}[t]
\centering
\caption{\detector{}'s Accuracy@1 with Various LLMs}
\label{tab: baseline cmp}
\small
\begin{threeparttable}
\begin{tabular}{lccccc}
\toprule
Approach     & Java    & JS     & Python & Go  & Avg.   \\
\midrule
\multicolumn{5}{l}{\textit{Ranking-based Non-LLMs:}}               \\
FastXML       & 0.292   & 0.078  & 0.491  & 0.277 & 0.285 \\
LightXML      & 0.450   & 0.146  & 0.529  & 0.494 & 0.405 \\
Chronos       & 0.516   & 0.447  & 0.550  & 0.710 & 0.556 \\
VulLibMiner   & 0.669   & 0.742  & 0.825  & 0.647 & 0.721 \\
\midrule
\multicolumn{5}{l}{\textit{Commercial LLMs:}} &              \\
ChatGPT       & 0.758   & 0.732  & 0.915  & 0.646  & 0.763\\
GPT4          & \textbf{0.783}   & 0.768  & 0.868  & 0.712 & 0.783 \\
\midrule
\multicolumn{5}{l}{\textit{Few-Shot ICL on Open-Source LLMs:}} & \\
LLaMa-7B      & 0.002   & 0.237  & 0.036  & 0.000 & 0.069 \\
LLaMa-13B     & 0.122   & 0.238  & 0.049  & 0.048 & 0.114 \\
Vicuna-7B     & 0.110   & 0.495  & 0.694  & 0.428 & 0.432 \\
Vicuna-13B    & 0.186   & 0.513  & 0.527  & 0.394 & 0.405 \\
\midrule
\multicolumn{5}{l}{\textit{Full SFT on Open-Source LLMs:}} &   \\
LLaMa-7B      & 0.710  & \textbf{0.773}  & 0.924  & 0.716 & 0.781 \\
LLaMa-13B     & 0.720  & 0.765  & 0.904  & 0.775 & 0.791 \\
Vicuna-7B     & 0.697  & 0.768  & 0.929  & 0.782 & 0.794 \\
Vicuna-13B    & 0.710  & \textbf{0.773}  & \textbf{0.935}  & \textbf{0.804} & \textbf{0.806} \\
\bottomrule
\end{tabular}
\end{threeparttable}
\vspace{-0.3cm}
\end{table}

\noindent \textbf{Overall Performance: Existing Work vs. \detector{}}. In Table~\ref{tab: baseline cmp}, Table~\ref{tab: baseline cmp: recall} and Table~\ref{tab: baseline cmp: F1} (Appendix), we compare the performance of \detector{} and baselines. 
From these tables we can observe that for all programming languages, \detector{} achieves substantially higher accuracies compared to existing work. As a result, by leveraging LLMs, \detector{} can effectively generate the names of affected packages with high accuracies. 

Overall, \detector{} using supervised fine-tuning on the Vicuna-13B model has the best performance. Fine-tuning Vicuna-13B even outperforms the larger ChatGPT and GPT4 models on all datasets besides Java. As a result, the knowledge gap of LLMs can be effectively bridged by leveraging supervised fine-tuning. In Table~\ref{tab: p value} of Appendix, we further report statistical significance tests~\cite{ttest} between the overall best-performing generative approach (i.e., \detector{} using Vicuna-13B SFT) and the best-performing existing work (i.e., VulLibMiner~\cite{vullibminer}). The p-values in all tests are smaller than $1e\textit{-}5$.

\noindent \textbf{When Is \detector{} More Advantageous?} From Table~\ref{tab: baseline cmp} observe that the gap between \detector{} Vicuna-13B SFT
and the best-performing existing approach for each programming language are 0.041, 0.031, 0.11, and 0.157. By comparing with the data statistics in Table~\ref{tab: dataset info}, we can see that this gap is highly correlated with \emph{\#Unique packages in the dataset} and \emph{\#Total packages in the ecosystem}. In general, \detector{} is less advantageous when the output package name is longer and has a larger token space.

\begin{figure}[t]
\centering
\includegraphics[width=1\linewidth]{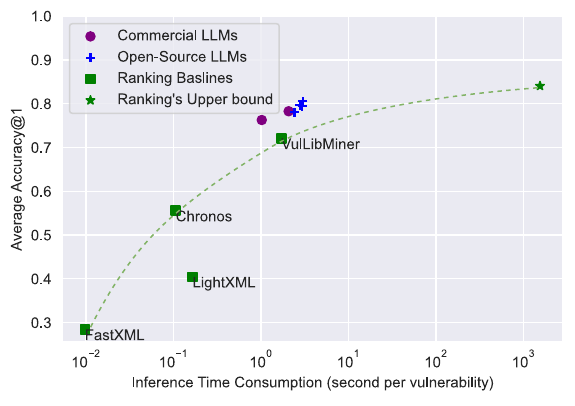}
\caption{Trade-Offs between Efficiency and Accuracy}
\label{fig: efficiency accuracy}
\vspace{-0.3cm}
\end{figure}



\noindent \textbf{Efficiency of Existing Work vs. \detector{}}. In Figure~\ref{fig: efficiency accuracy}, we visualize the actual computational cost and accuracy of each method in Table~\ref{tab: baseline cmp}. We further mark the upper bound of the ranking-based approach using LLMs for comparison. The Accuracy@1 is upper-bounded by the recall@512 of TF-IDF, i.e., the best possible Accuracy@1; while the time cost is upper-bounded by the time cost of invoking the 13B model 512 times (10 mins).

We can observe the \detector{} achieves a sweet spot in the effectiveness and efficiency trade-off. When compared with existing work, \detector{} achieves a better Accuracy@1 while the time cost is comparable to the best-performed existing work~\cite{vullibminer}. When compared with the upper bound, \detector{} achieves a slightly lower Accuracy@1 while consuming less than 1/100 time and computation resources.



\subsection{Ablation Studies on \detector{}}
In this subsection, we conduct ablation studies on the three components of \detector{}: supervised fine-tuning, RAG, and local search.

\noindent \textbf{SFT's Improvement}. By comparing the results of in-context learning vs supervised fine-tuning in Table~\ref{tab: baseline cmp}, we can see that SFT outperforms ICL by a larger margin.
This result indicates that for the 7B and 13B models, supervised fine-tuning on the full training data is essential in bridging the models' knowledge gap. 


\noindent \textbf{RAG's Overall Improvement:}
Table~\ref{tab: rag improvement} shows the improvement of our RAG technique in Accuracy@1.
Specifically, it improves the Accuracy@1 by 9.3\%, 1.8\%, 8.9\%, and 15.7\% on each programming language, respectively.
These improvements indicate that our RAG technique is effective in helping generate the names of vulnerable packages. 
We further report paired t-test results for Table~\ref{tab: rag improvement} in Table~\ref{tab: rag p value} of Appendix. 

Table~\ref{tab: rag improvement} also indicates that RAG's improvement in commercial LLMs is higher than that of open-source LLMs.
Especially in Go vulnerabilities, our RAG technique improves the Accuracy@1 by 57.6\% and 31.0\% on ChatGPT and GPT4.
The main reason is that both ChatGPT and GPT4 do not have sufficient domain knowledge about Go packages as they are relatively newer than packages of other programming languages~\cite{chatgpt_golang}. 



\begin{table}[t]
\centering
\small
\caption{RAG's Improvement ($Accuracy@1_{RAG} - Accuracy@1_{Raw}$)}
\label{tab: rag improvement}
\begin{tabular}{lrrrr}
\toprule
\multicolumn{1}{c}{\multirow{1}{*}{Language}} & \multicolumn{1}{c}{Java}                        &        \multicolumn{1}{c}{JS}                 &          \multicolumn{1}{c}{Python}               &     \multicolumn{1}{c}{Go}                    \\
\midrule
\multicolumn{3}{l}{\textit{Commercial LLMs:}}                          &                         &                         \\
ChatGPT                                       & 16.1\% $\uparrow$       & 3.6\% $\uparrow$       & 38.8\% $\uparrow$       & 57.6\% $\uparrow$                \\
GPT4                                          & 12.1\% $\uparrow$       &    0.9\% $\uparrow$                     &      0.9\% $\uparrow$                   &        31.0\% $\uparrow$                 \\ 
\midrule
\multicolumn{5}{l}{\textit{Full SFT on Open-Source LLMs:}}                                                                          \\
LLaMa-7B                                      & 2.2\% $\uparrow$       & 2.2\% $\uparrow$       & 3.1\% $\uparrow$       & 1.8\% $\downarrow$     \\
LLaMa-13B                                     & 3.3\% $\uparrow$       & 0.4\% $\uparrow$       & 2.0\% $\uparrow$       & 3.7\% $\uparrow$       \\
Vicuna-7B                                     & 13.6\% $\uparrow$      & 2.3\% $\uparrow$       & 3.8\% $\uparrow$       & 3.7\% $\uparrow$       \\
Vicuna-13B                                    & 8.7\% $\uparrow$      & 1.2\% $\uparrow$       & 4.9\% $\uparrow$       & 0.0\% -                \\
\midrule
Average                                      & 9.3\% $\uparrow$      & 1.8\% $\uparrow$       & 8.9\% $\uparrow$       & 15.7\% $\uparrow$         \\
\bottomrule
\end{tabular}
\end{table}

\noindent \textbf{RAG Improvement vs. k/Retrieval Algorithm Choice}. We evaluate whether $k$ and the choice of retrieval algorithm affect the end-to-end effectiveness of \detector{}. Specifically, we focus on Java vulnerabilities (as Java package names are the most difficult to generate). 
The result can be found in Table~\ref{tab: various rag} in our Appendix.

For $k$, we conduct an Analysis of Variance (ANOVA)~\cite{anova} among the Accuracy@1 of six representative numbers of RAG packages (ranging from 1 to 20). Although $k=20$ has a slightly higher accuracy than $k=1$ for both TF-IDF and BERT, this difference is not significant. In fact, the paired t-test results show that there is no significant difference among the Accuracy@1 of different $k$ values ($p=0.814$ for TF-IDF and $p=0.985$ for BERT). 

As for the retrieval algorithm, we observe that Accuracy@1 with TF-IDF results is quite similar to that of non-RAG inputs, and the Accuracy@1 with BERT results is substantially higher than that of non-RAG/TF-IDF results. As a result, it is essential to use BERT-retrieved results in RAG.

\noindent \textbf{Local Search's Improvement}. 
Table~\ref{tab: post processing} shows the end-to-end 
improvement in Accuracy@1 of \detector{} before and after local search.
Our local search technique improves the Accuracy@1 by 3.43\%, 1.02\%, 1.57\%, and 6.20\% on each programming language. We further report paired t-test results for Table~\ref{tab: post processing} in Table~\ref{tab: local search p value} of Appendix. 

We make the following observations. First, local search is more effective on commercial LLMs (an average improvement of 4.58\%) than fine-tuned open-source LLMs (an average improvement of 2.29\%).
Since commercial LLMs are not fine-tuned, local search plays an important role in improving the effectiveness of generation. Second, local search is more effective on Java and Go than JS and Python. The reason is that since Java and Go packages are longer (8-14 tokens), LLMs are more prone to generating partially correct, non-existing outputs (i.e., Type 2 error in Table~\ref{tab: fault case study}).  Local search can effectively reduce this type of error. 

\begin{table}[t]
\centering
\small
\caption{Local Search's Improvement ($Accuracy@1_{Search} - Accuracy@1_{Raw}$)}
\label{tab: post processing}
\begin{tabular}{lrrrr}
\toprule
\multicolumn{1}{c}{\multirow{1}{*}{Language}} & \multicolumn{1}{c}{Java}                        &        \multicolumn{1}{c}{JS}                 &          \multicolumn{1}{c}{Python}               &     \multicolumn{1}{c}{Go}                    \\
\midrule
\multicolumn{3}{l}{\textit{Commercial LLMs:}}                          &                         &                         \\
ChatGPT                                       & 4.1\% $\uparrow$       & 1.2\% $\uparrow$       & 0.7\% $\uparrow$       & 11.1\% $\uparrow$                \\
GPT4                                          & 5.3\% $\uparrow$       &    2.5\% $\uparrow$                     &      2.9\% $\uparrow$                   &        8.8\% $\uparrow$                 \\ 
\midrule
\multicolumn{5}{l}{\textit{Full SFT on Open-Source LLMs:}}                                           \\
LLaMa-7B                                      & 2.9\% $\uparrow$       & 0.9\% $\uparrow$       & 2.2\% $\uparrow$       & 7.0\% $\uparrow$     \\
LLaMa-13B                                     & 3.3\% $\uparrow$       & 0.3\% $\uparrow$       & 0.7\% $\uparrow$       & 4.1\% $\uparrow$       \\
Vicuna-7B                                     & 3.9\% $\uparrow$      & 0.9\% $\uparrow$       & 1.8\% $\uparrow$       & 3.3\% $\uparrow$       \\
Vicuna-13B                                    & 1.1\% $\uparrow$      & 0.3\% $\uparrow$       & 1.1\% $\uparrow$       & 2.9\% $\uparrow$       \\
\midrule
Average                                      & 3.4\% $\uparrow$      & 1.0\% $\uparrow$       & 1.4\% $\uparrow$       & 6.2\% $\uparrow$         \\
\bottomrule
\end{tabular}
\end{table}

\subsection{Evaluating \detector{} Performance in Real World Setting}~\label{sec: real world}

\noindent To examine \detector{}'s performance in the real-world setting, for each programming language, we randomly sample and report a subset of <vulnerability, affected package> pairs that are not listed in GitHub Advisory (Java: 25, JS: 14, Python: 11, Go: 10). 
We use \detector{} to generate the package names and submit the generated names (\detector{} with Vicuna-13B) to GitHub Advisory. 

At the time of the writing,  the results are summarized below. \textbf{Java}: 21 of them have been accepted and merged into GitHub Advisory. Among the remaining 4 packages, 2 of them are considered non-vulnerabilities, and 2 of them are considered incorrect affected packages. \textbf{JS, Python, and Go}: 13 of them have been accepted and merged into GitHub Advisory (2 JS, 8 Python, 3 Go), and the remaining 20 have been submitted and are waiting for review. The details of these packages are listed in the Appendix (Table~\ref{tab: github issue}). 

This result highlights the real-world performance of \detector{} in automatically identifying affected package names.

%% file: related.tex
\section{Related Work}

\textbf{Vulnerable Package/Version Identification}. Numerous existing works have proposed methods to improve the accuracy of affected package identification. Multiple existing works model this problem as a named entity recognition (NER) problem, i.e., extracting the subset of description about the package~\cite{viem, anwar2021cleaning, jo2022vulcan, kuehn2021ovana,yang2021few} or version~\cite{viem, Atvhunter, Libid, libscout, libpecker, libdb, gorla2014checking, wu2023understanding}. The NER approach works well for the software version identification since many version numbers are already in the description~\cite{viem}. On the other hand, the package names are often only partially mentioned (e.g., \href{https://github.com/advisories/GHSA-264w-xrr7-6qqg}{CVE-2020-2167} in Table~\ref{tab: fault case study}), therefore the NER approaches are less effective~\cite{chronos}. Another branch of work models the package identification problem as extreme multi-label learning (XML) where each package is a class~\cite{fastxml, lightxml, chronos}. However, these methods are limited to less than 3k classes (the labels in their dataset). Finally, \cite{vullibminer} leverages the re-ranking approach using BERT; however, there still exists a gap between their method's accuracy and the best possible performance (Table~\ref{tab: baseline cmp}).

\noindent\textbf{Retrieval vs Generation}. Existing work has investigated scenarios of replacing the retrieval with generation. For example, Yu et al.~\cite{yu2022generate} leverages LLM to generate the context documents for question answering, rather than retrieving them from a text corpus. Their experiment shows that the generative approach has a comparable performance to the retrieval approach on the QA task. 
However, since our task requires us to generate the exact package name, their conclusion is not directly transferrable to our task.



\noindent \textbf{Retrieval-Augmented Generation}
Retrieval-augmented generation (RAG)~\cite{lewis2020retrieval,
mao2020generation, liu2020retrieval, cai2022recent} is a widely used technique and has shown its effectiveness in various generation tasks, e.g., code generation or question answering.
Specifically, RAG enhances the performance of a generative model by incorporating knowledge from a database so that LLMs can extract and comprehend correct domain knowledge from the RAG inputs.

\noindent\textbf{Reducing Hallucination}. 
In Section~\ref{sec: empirical_study}, we show that ChatGPT's raw output package name may not exist. This phenomenon is similar to hallucination~\cite{ji2023survey, tonmoy2024comprehensive}, which occurs in various LLM-related tasks.
Among hallucination reduction approaches, post-processing~\cite{madaan2023self, kang2023large} is a widely used one.
For example, in code-related tasks, existing work~\cite{jin2023inferfix, chen2023codet, zhang2023self, huynh2022hierarchynet} adopts post-processing techniques to reduce/rerank programs generated by LLMs, e.g., using deep-learning models, test cases, or compilers to determine whether a generated program is correct and remove incorrect programs.
However, such techniques cannot be directly adopted in our task because validating the generated names of affected packages is relatively difficult. 
It requires a Proof-of-Chain (PoC)~\cite{mosakheil2018security}, which is often unavailable due to security concerns.
Therefore, we design our local search algorithm focusing on Type 2 errors in Table~\ref{tab: fault case study}.

%% file: conclusion.tex
\section{Conclusion}\label{sec:conclusion}
In this paper, we have proposed \detector{}, the first framework for identifying vulnerable packages using LLM generation. \detector{} conducts retrieval-augmented generation, supervised fine-tuning, and a local search technique to improve the generation. \detector{} is highly effective, achieving an accuracy of 0.806 while the best SOTA approaches achieve only 0.721.
\detector{} has shown high value to security practice. 
We have submitted 60 pairs of <vulnerability, affected package> to GitHub advisory. 34 of them have been accepted and merged, and 20 are pending approval.

%% file: limitations.tex
\newpage

\section{Limitation}~\label{sec: limitation}
Our work has several limitations, which we plan to address in our future work:

\noindent \textbf{Challenges in Generating Long and Complex Package Names}. 
As discussed in Section~\ref{sec: trade off}, the effectiveness of \detector{} depends on the token length and the number of unique packages.
Table~\ref{tab: baseline cmp} shows Java is more challenging than others while having the highest token length and unique packages (Table~\ref{tab: dataset info}). 
To improve the generation of complicated languages such as Java, we plan to further enhance the knowledge of LLM using techniques such as constrained decoding~\cite{post-vilar-2018-fast}. We leave this as our future work.
In particular, it may pose further challenges to generate packages that are exceptionally long. To understand the distribution of token lengths, we report the quantile statistics of the token lengths in Table~\ref{tab: quantile} of Appendix. Table~\ref{tab: quantile} shows that the majority of the package names are shorter than 20 tokens, therefore, exceptionally long package names are very rare. 

\noindent \textbf{Challenges in Generating Package Names with Limited Ecosystem Knowledge}. 
Though \detector{} has demonstrated its effectiveness in four widely-used programming languages, some other programming languages, e.g., C/C++, do not have a commonly used ecosystem that maintains all its packages.
Thus, it is difficult to generate/retrieve the affected packages of C/C++ vulnerabilities as we do not have specific ranges during the RAG step of \detector{}.
Exploring how to generate RAG results without a commonly used ecosystem (e.g., Maven or Pypi) or collecting other useful information for RAG is the future work of this paper.


\section{Ethical Consideration}

\noindent \textbf{License/Copyright.}
\detector{} utilizes open-source data from GitHub Advisory, along with four third-party package ecosystems.
We refer users to the original licenses accompanying the resources of these data.

\noindent \textbf{Intended Use.}
\detector{} is designed as an automatic tool to assist maintainers of vulnerability databases, e.g., GitHub Advisory.
Specifically, \detector{} helps generate the names of affected packages to complement the missing data of these databases.
The usage of \detector{} is also illustrated in Section~\ref{sec:approach} and our intended usage of \detector{} is consistent with that of GitHub Advisory~\cite{githubAD}.

\noindent \textbf{Potential Misuse.}
Similar to existing open-source LLMs, one potential misuse of \detector{} is generating harmful content.
Considering that we use open-source vulnerability data for LLM fine-tuning, the LLM might view harmful content during this step.
To avoid harmful content, we use only reviewed vulnerability data in GitHub Advisory, so such misuse will unlikely happen. 
Overall, the scientific and social benefits of the research arguably outweigh the small risk of their misuse.

%% file: appendix.tex
\section{Appendix}~\label{sec: appendix}


\subsection{Our Local Search Algorithm}~\label{sec: local search algorithm}

\begin{algorithm}[h]
\small
	\SetKwData{eval}{\textbf{eval}}
    \SetKwData{argmin}{argmin}
	\SetKwFunction{compatible}{compatible}	
	\SetKwInOut{Input}{Input}\SetKwInOut{Output}{Output}
 
    \Input{$rawName$, a generated package name}
    \Output{$vulnNames$, names of affected packages.}
    \BlankLine
            \BlankLine
        \tcp{Pre-process on name list}
        $nameDict, suffixes = \gets \{\}, \emptyset{}$\;
        \For{$name \in nameList$}{
            $prefix, suffix \gets name.split(\mbox{``/:''})$\;
            $nameDict[suffix].add(prefix)$\;
        }


        \BlankLine
        \tcp{Search the closest prefix/suffix}
        $prefix, suffix \gets rawName.split(\mbox{``/:''})$\;
        $edit.weight \gets (W_{insert}, W_{delete}, W_{replace})$\;
        $suffix^{\prime} \gets \mathop{\argmin}\limits_{s \in suffixes}{edit(suffix, s)}$\;
        $prefixes \gets nameDict[suffix^{\prime}]$\;
        \If{$prefixes.isEmpty()$}{
            \Return{$\{suffix^{\prime}\}$}\;
        }
        \Else{
            $prefix^{\prime} \gets \mathop{\argmin}\limits_{p \in prefixes}{edit(prefix, p)}$\;
            \Return{$\{prefix^{\prime}\}: \{suffix^{\prime}\}$}\;
            }
        
\caption{Local Search}\label{alg: local search}
\end{algorithm}

The pseudocode of our local search algorithm is shown in Algorithm~\ref{alg: local search}.
The input of this algorithm includes one package name generated by LLM together with the name list of existing libraries under the same ecosystem. 
The output of this algorithm is the name of one existing package that is the closest to the generated package name.

In Lines 1-4, we pre-process the name list of candidate packages.
Now that we divide a package name into its prefix and suffix, we first construct the dictionary $nameDict$ that maps a suffix into its corresponding prefix.


In Lines 5-13, we search for the closest package name of the input package name, $rawName$.
In Line 7, we use its suffix, $suffix$ to find its closest and existing suffix, $suffix^{\prime}$.
Then in Lines 8-13, we first determine whether it contains a corresponding prefix.
If it has no prefix (e.g., a Python package), we directly return the closest suffix.
Otherwise, we find its closest prefix, $prefix^{\prime}$, from all prefixs that correspond to $suffix^{\prime}$.
Additionally, in Line 6, we manually set the weight used in calculating edit distances because LLMs change the package names in terms of tokens instead of characters.
Thus, the weight of inserting one character should be smaller than that of deleting and replacing one, and we set the empirical weights as follows, $W_{insert} = 1, W_{delete} = 4, W_{replace} = 4$.


\begin{table}[t]
\centering
\caption{The Quantile Statistics of the Number of Tokens in the Package Names}
\label{tab: quantile}
\small
\begin{tabular}{lllll}
\toprule
{quantile} & {0.5} & {0.75} & {0.95} & {1} \\
\midrule
Maven             & 13           & 15            & 20            & 40         \\
npm               & 4            & 6             & 9             & 16         \\
pypi              & 3            & 5             & 8             & 15         \\
go                & 13           & 15            & 18            & 34         \\
\bottomrule
\end{tabular}
\end{table}

\subsection{The Performance Comparison between \detector{} and Baselines}

The performance comparison (Recall, F1) between \detector{} and baselines can be found in Table~\ref{tab: baseline cmp: recall} and Table~\ref{tab: baseline cmp: F1}.

\begin{table}[h]
\centering
\caption{\detector{}'s Recall@1 with Various LLMs}
\label{tab: baseline cmp: recall}
\small
\begin{threeparttable}
\begin{tabular}{lccccc}
\toprule
Approach     & Java    & JS     & Python & Go  & Avg.   \\
\midrule
\multicolumn{5}{l}{\textit{Ranking-based Non-LLMs:}}               \\
Chronos       & 0.4	& 0.412	& 0.286	& 0.605	& 0.426 \\
VulLibMiner   & 0.520	&0.709	&0.499	&0.544	&0.568\\
\midrule
\multicolumn{5}{l}{\textit{Commercial LLMs:}} &              \\
ChatGPT       & 0.573&	0.698	&0.580&	0.544	&0.599\\
GPT4          & \textbf{0.596}	&0.714	&0.542	&0.580&	0.608 \\
\midrule
\multicolumn{5}{l}{\textit{Full SFT on Open-Source LLMs:}} &   \\
LLaMa-7B      & 0.548	&0.734&	0.593	&0.625&	0.625 \\
LLaMa-13B     & 0.554	&0.73	&0.601	&0.656	&0.635 \\
Vicuna-7B     & 0.543	&0.731	&0.601&	0.66	&0.634 \\
Vicuna-13B    & 0.552	&\textbf{0.736}&	\textbf{0.621}	&\textbf{0.688}&	\textbf{0.649 }\\
\bottomrule
\end{tabular}
\end{threeparttable}
\vspace{-0.4cm}
\end{table}

\begin{table}[h]
\centering
\caption{\detector{}'s F1@1 with Various LLMs}
\label{tab: baseline cmp: F1}
\small
\begin{threeparttable}
\begin{tabular}{lccccc}
\toprule
Approach     & Java    & JS     & Python & Go  & Avg.   \\
\midrule
\multicolumn{5}{l}{\textit{Ranking-based Non-LLMs:}}               \\
Chronos       & 0.451 & 	0.429 & 	0.376	 & 0.653	 & 0.482
 \\
VulLibMiner   & 0.585	 & 0.725	 & 0.622 & 	0.591 & 	0.635
\\
\midrule
\multicolumn{5}{l}{\textit{Commercial LLMs:}} &              \\
ChatGPT       & 0.653 & 	0.715 & 	0.710	 & 0.591 & 	0.671
\\
GPT4          & \textbf{0.677}	 & 0.740	 & 0.667	 & 0.639 & 	0.684
\\
\midrule
\multicolumn{5}{l}{\textit{Full SFT on Open-Source LLMs:}} &   \\
LLaMa-7B      & 0.619 & 	0.753	 & 0.722 & 	0.667 & 	0.694
 \\
LLaMa-13B     & 0.626 & 	0.747 & 	0.722 & 	0.711	 & 0.705
 \\
Vicuna-7B     & 0.610	 & 0.749	 & 0.730 & 	0.716	 & 0.705
 \\
Vicuna-13B    & 0.621	 & \textbf{0.754}	 & \textbf{0.746}	 & \textbf{0.741}	 & \textbf{0.719}
\\
\bottomrule
\end{tabular}
\end{threeparttable}
\vspace{-0.4cm}
\end{table}

\begin{table*}[t]
\centering
\small
\caption{Accuracy@1 with Various RAG Inputs in Generating the Names of Java Affected Packages}
\label{tab: various rag}
\begin{tabular}{llcccccccccccc}
\toprule
IR Model:      & None  & \multicolumn{6}{c}{TF-IDF Results}            & \multicolumn{6}{c}{BERT Results}              \\
\cmidrule(lr){3-8}\cmidrule(lr){9-14}
\multicolumn{2}{l}{\#RAG packages:} & 1                & 2               & 3     & 5       & 10      & 20      & 1               & 2              & 3     & 5       & 10      & 20      \\
\midrule
\multicolumn{3}{l}{\textit{Commercial LLMs:}}                    &                 &       &         &         &         &                 &                &       &         &         &         \\
ChatGPT    & 0.597 & 0.523 & 0.498 & 0.508 & 0.552 & 0.540 & 0.567 & 0.758 & 0.743 & 0.722 & 0.718 & 0.715 & 0.710 \\
GPT4       & 0.676 & 0.619 & 0.559 & 0.588 & 0.619 & 0.626 & 0.638 & 0.783 & 0.773 & 0.784 & 0.792 & 0.797 & 0.792 \\
\midrule
\multicolumn{5}{l}{\textit{Full SFT on Open-Source LLMs:}}                        &         &         &         &                 &                &       &         &         &         \\
LLaMa-7B   & 0.688 & 0.692 & 0.697 & 0.701 & 0.563 & 0.591 & 0.609 & 0.710 & 0.701 & 0.710 & 0.678 & 0.665 & 0.683 \\
LLaMa-13B  & 0.687 & 0.688 & 0.687 & 0.696 & 0.653 & 0.635 & 0.623 & 0.720 & 0.702 & 0.701 & 0.701 & 0.704 & 0.703 \\
Vicuna-7B  & 0.561 & 0.596 & 0.398 & 0.404 & 0.441 & 0.439 & 0.421 & 0.697 & 0.701 & 0.683 & 0.685 & 0.706 & 0.683 \\
Vicuna-13B & 0.623 & 0.609 & 0.450 & 0.418 & 0.650 & 0.655 & 0.680 & 0.710 & 0.712 & 0.701 & 0.722 & 0.719 & 0.720 \\
\bottomrule
\end{tabular}
\vspace{-0.4cm}
\end{table*}

\subsection{Statistical Significance Test between \detector{} and Baselines\label{sec: appendix tables}}

\textbf{\detector{} vs the Best Baseline}. Table~\ref{tab: p value} shows the p-values of \detector{} between the best-performing generative approach (i.e., \detector{} using Vicuna-13B SFT) and the best-performing existing work (i.e., VulLibMiner~\cite{vullibminer}).
From this table, we can observe that the p-values of all tests are smaller than $1e\textit{-}5$, indicating the significant improvement of \detector{}'s effectiveness.

\begin{table}[t]
\centering
\caption{Parameters Used in Fine-Tuning LLMs}
\label{tab: hyper parameter}
\small
\begin{tabular}{l|l}
\toprule
\multicolumn{2}{l}{\textit{Supervised Fine-Tuning Parameters:}}\\
\cmidrule(lr){0-1}
Train Batch Size : 4 & Learning Rate : 2e-5      \\
Evaluation Batch Size : 4 & Weight Decay  : 0.00        \\
Learning Rate schedule : Cosine    & Warmup Ratio  : 0.03         \\
Max Sequence Length: 512 & Use Lora: True \\
\midrule
\multicolumn{2}{l}{\textit{In-Context Learning Parameters:}}\\
\cmidrule(lr){0-1}
Max Sequence Length: 512 & \#Shots : 3 \\
\bottomrule
\end{tabular}
\end{table}

\begin{table}[h]
\centering
\caption{The P-Values of \detector{} (Compared with VulLibMiner)}
\label{tab: p value}
\small
\begin{threeparttable}
\begin{tabular}{llllll}
\toprule
Approach     & Java    & JS     & Python & Go  & Avg.   \\
\midrule
\multicolumn{5}{l}{\textit{Commercial LLMs:}} &              \\
ChatGPT       & 2e-13   & 8e-3  & 1e-10  & 3e-1  & 7e-2\\
GPT4          & 1e-18  & 5e-5  & 1e-5  & 3e-5 & 2e-5 \\
\midrule
\multicolumn{5}{l}{\textit{Full SFT on Open-Source LLMs:}} & \\
LLaMa-7B      & 7e-7   & 1e-5  & 1e-11  & 9e-6 & 5e-6 \\
LLaMa-13B     & 5e-8   & 1e-4  & 1e-9  & 1e-9 & 4e-5 \\
Vicuna-7B     & 5e-5   & 6e-5  & 1e-12  & 5e-10 & 3e-5 \\
Vicuna-13B    & 7e-7   & 1e-5  & 6e-13  & 1e-11 & 3e-6 \\
\midrule
Average       &  1e-5 & 1e-3 & 2e-6 & 5e-2 & 1e-2\\
\bottomrule
\end{tabular}
\end{threeparttable}
\end{table}

\noindent \textbf{Significance of Using Local Search}. Table~\ref{tab: local search p value} shows the p-values of the paired t-tests for each number in Table~\ref{tab: post processing}, i.e., before and after local search. We highlight the p-values that are larger than 0.05 in Table~\ref{tab: local search p value}. We can observe that local search's improvement is less significant in JS and Python and more significant in Java and Go.

\begin{table}[h]
\centering
\caption{The P-Values of \detector{} Before and After Local Search}
\label{tab: local search p value}
\small
\begin{threeparttable}
\begin{tabular}{llllll}
\toprule
Language   & Java & JS            & Python        & Go   & Avg. \\
\midrule
\multicolumn{6}{l}{\textit{Commercial LLMs:}}                              \\
ChatGPT    & 1e-6 & 8e-3          & \textbf{0.08} & 1e-8 & 2e-2  \\
GPT4       & 3e-8 & 6e-5          & 2e-4          & 1e-6 & 6e-5  \\
\midrule
\multicolumn{6}{l}{\textit{Fine-Tuned Open-Source LLMs:}}                  \\
LLaMa-7B   & 3e-5 & 2e-2          & 2e-3          & 9e-6 & 5e-3  \\
LLaMa-13B  & 2e-5 & \textbf{0.16} & \textbf{0.08} & 4e-4 & 6e-2  \\
Vicuna-7B  & 2e-6 & 2e-2          & 4e-3          & 2e-3 & 6e-3  \\
Vicuna-13B & 8e-3 & \textbf{0.32} & 2e-2          & 8e-3 & 3e-2  \\
\midrule
Average    & 1e-3 & \textbf{0.09} & 3e-2          & 1e-3 & 3e-2  \\
\bottomrule
\end{tabular}
\end{threeparttable}
\end{table}

\noindent \textbf{Significance of Using RAG}.
Table~\ref{tab: rag p value} shows the p-values of the paired t-tests for each number in Table~\ref{tab: rag improvement}, i.e., before and after RAG.
We highlight the p-values that are larger than 0.05 in Table~\ref{tab: rag p value}. We can observe that RAG significantly improves the effectiveness of \detector{} in most combinations of LLMs and programming languages.

\begin{table}[h]
\centering
\caption{The P-Values of \detector{} Before and After RAG}
\label{tab: rag p value}
\small
\begin{tabular}{llllll}
\toprule
Language    & Java   & JS     & Python & Go     & Avg. \\
\midrule
\multicolumn{5}{l}{Commercial LLMs}             &         \\
ChatGPT     & 3e-23 & 1e-06 & 3e-49 & 1e-52 & 3e-07  \\
GPT4        & 1e-17 & 2e-02 & 4e-02 & 2e-23 & 2e-02  \\
\midrule
\multicolumn{5}{l}{Fine-Tuned Open-Source LLMs} &         \\
LLaMa-7B    & 3e-04 & 2e-04 & 3e-04 & 4e-02 & 1e-02  \\
LLaMa-13B   & 2e-05 & \textbf{0.16} & 5e-03 & 8e-04 & 4e-02  \\
Vicuna-7B   & 5e-13 & 8e-03 & 2e-06 & \textbf{nan}   & 2e-03  \\
Vicuna-13B  & 8e-03 & 3e-01 & 2e-02 & 8e-03 & 9e-02  \\
\midrule
Average     & 1e-03 & 8e-02 & 1e-02 & 1e-02 & 3e-02  \\
\bottomrule
\end{tabular}
\end{table}

\begin{table*}[t]
\centering
\caption{Status of Submitted <Vulnerability, Affected Package> Pairs}
\label{tab: github issue}
\small
\begin{threeparttable}
\begin{tabular}{lllc}
\toprule
CVE ID & Language & \detector{}'s Output & Status \\
\midrule
\href{https://github.com/advisories/GHSA-hr3v-8cp3-68rf}{CVE-2021-41803} & Go & github.com/hashicorp/consul & submitted \\
\href{https://github.com/advisories/GHSA-hhvx-8755-4cvw}{CVE-2023-1296} & Go & github.com/hashicorp/nomad & submitted \\
\href{https://github.com/advisories/GHSA-33j2-92xf-fwm3}{CVE-2023-2197} & Go & github.com/hashicorp/vault & submitted \\
\href{https://github.com/advisories/GHSA-rpvr-38xv-xvxq}{CVE-2023-3072} & Go & github.com/hashicorp/nomad & Merged \\
\href{https://github.com/advisories/GHSA-fhmj-jv7w-vvg2}{CVE-2023-3114} & Go & github.com/hashicorp/terraform & submitted \\
\href{https://github.com/advisories/GHSA-9jfx-84v9-2rr2}{CVE-2023-3299} & Go & github.com/hashicorp/nomad & Merged \\
\href{https://github.com/advisories/GHSA-v5fm-hr72-27hx}{CVE-2023-3300} & Go & github.com/hashicorp/nomad & submitted \\
\href{https://github.com/advisories/GHSA-9rhf-q362-77mx}{CVE-2023-3518} & Go & github.com/hashicorp/consul & Merged \\
\href{https://github.com/advisories/GHSA-7j85-mwfj-2gr8}{CVE-2023-3774} & Go & github.com/hashicorp/vault & submitted \\
\href{https://github.com/advisories/GHSA-37gg-8xjr-m6x4}{CVE-2023-3775} & Go & github.com/hashicorp/vault & submitted \\
\href{https://github.com/advisories/GHSA-w97x-8w5v-6mh4}{CVE-2007-2379} & JS & jquery & submitted \\
\href{https://github.com/advisories/GHSA-jjg9-mf63-vqrp}{CVE-2012-5881} & JS & yui2 & Merged \\
\href{https://github.com/advisories/GHSA-3jcq-cwr7-6332}{CVE-2013-2022} & JS & jplayer & submitted \\
\href{https://github.com/advisories/GHSA-mjh3-g7qw-vgfv}{CVE-2013-4383} & JS & jquery-countdown & submitted \\
\href{https://github.com/advisories/GHSA-xc36-3p8q-x8x7}{CVE-2013-6837} & JS & types/jquery.prettyphoto & submitted \\
\href{https://github.com/advisories/GHSA-q44p-q588-242q}{CVE-2014-6071} & JS & jquery & Non-Vuln \\
\href{https://github.com/advisories/GHSA-q44p-q588-242q}{CVE-2014-6071} & JS & jquery & submitted \\
\href{https://github.com/advisories/GHSA-jcxc-mh25-387r}{CVE-2018-7747} & JS & calderajs/forms & Incorrect \\
\href{https://github.com/advisories/GHSA-pfm2-mqwj-ggm5}{CVE-2020-10960} & JS & mediawiki & Merged \\
\href{https://github.com/advisories/GHSA-v63q-hgqc-qvpg}{CVE-2021-32821} & JS & mootools & submitted \\
\href{https://github.com/advisories/GHSA-8q87-cc79-vwjj}{CVE-2021-36713} & JS & datatables & submitted \\
\href{https://github.com/advisories/GHSA-43x9-7hfv-mxrf}{CVE-2021-37504} & JS & jquery-file-upload & submitted \\
\href{https://github.com/advisories/GHSA-9p4g-cjcf-q3x2}{CVE-2021-43956} & JS & fisheye & submitted \\
\href{https://github.com/advisories/GHSA-9p4g-cjcf-q3x2}{CVE-2021-43956} & JS & crucible & submitted \\
\href{https://github.com/advisories/GHSA-97gm-mcv6-cphm}{CVE-2010-5327} & Java & com.liferay.portal:portal-impl & Merged \\
\href{https://github.com/advisories/GHSA-97gm-mcv6-cphm}{CVE-2010-5327} & Java & com.liferay.portal:portal-service & Merged \\
\href{https://github.com/advisories/GHSA-ppg2-ww3w-hq84}{CVE-2012-3428} & Java & org.jboss.ironjacamar:ironjacamar-jdbc & Merged \\
\href{https://github.com/advisories/GHSA-428j-q447-47rw}{CVE-2013-1814} & Java & org.apache.rave:rave-core & Merged \\
\href{https://github.com/advisories/GHSA-428j-q447-47rw}{CVE-2013-1814} & Java & org.apache.rave:rave-portal-resources & Merged \\
\href{https://github.com/advisories/GHSA-428j-q447-47rw}{CVE-2013-1814} & Java & org.apache.rave:rave-web & Merged \\
\href{https://github.com/advisories/GHSA-wf5v-jhxj-q632}{CVE-2014-0095} & Java & org.apache.tomcat.embed:tomcat-embed-core & Merged \\
\href{https://github.com/advisories/GHSA-wf5v-jhxj-q632}{CVE-2014-0095} & Java & org.apache.tomcat:tomcat-coyote & Merged \\
\href{https://github.com/advisories/GHSA-c2fp-mpmm-cqxv}{CVE-2014-1202} & Java & com.smartbear.soapui:soapui & Merged \\
\href{https://github.com/advisories/GHSA-q79q-94j7-5mgg}{CVE-2014-9515} & Java & com.github.dozermapper:dozer-parent & Non-Vuln \\
\href{https://github.com/advisories/GHSA-9qhq-j4xm-cw48}{CVE-2015-3158} & Java & org.picketlink:picketlink-bindings-parent & Incorrect \\
\href{https://github.com/advisories/GHSA-qhxw-54m9-6wwc}{CVE-2017-1000397} & Java & org.jenkins-ci.main:maven-plugin & Merged \\
\href{https://github.com/advisories/GHSA-4px2-gqhv-mrc7}{CVE-2017-1000406} & Java & org.opendaylight.integration:distribution-karaf & Merged \\
\href{https://github.com/advisories/GHSA-j88v-q3vw-p9vr}{CVE-2017-3202} & Java & com.exadel.flamingo.flex:amf-serializer & Merged \\
\href{https://github.com/advisories/GHSA-f5ch-36rg-vfcc}{CVE-2017-7662} & Java & org.apache.cxf.fediz:fediz-oidc & Merged \\
\href{https://github.com/advisories/GHSA-38xm-xhvj-q2qf}{CVE-2018-1000057} & Java & org.jenkins-ci.plugins:credentials-binding & Merged \\
\href{https://github.com/advisories/GHSA-6w3h-vq7m-v3qf}{CVE-2018-1000191} & Java & com.synopsys.integration:synopsys-detect & Merged \\
\href{https://github.com/advisories/GHSA-4cj8-779h-r25h}{CVE-2018-1229} & Java & org.springframework.batch:spring-batch-admin-manager & Merged \\
\href{https://github.com/advisories/GHSA-q4q2-93pw-qwgf}{CVE-2018-1256} & Java & 
 io.pivotal.spring.cloud:spring-cloud-sso-connector & Merged \\
\href{https://github.com/advisories/GHSA-mjpc-qx7h-r8c9}{CVE-2018-3824} & Java & org.elasticsearch:elasticsearch & Merged \\
\href{https://github.com/advisories/GHSA-j3r9-f742-jp74}{CVE-2018-5653} & Java & wordpress/weblizar-pinterest-feeds & Incorrect \\
\href{https://github.com/advisories/GHSA-f8w9-66fp-3jgw}{CVE-2019-10475} & Java & org.jenkins-ci.plugins:build-metrics & Merged \\
\href{https://github.com/advisories/GHSA-h755-h99p-9ffv}{CVE-2019-5312} & Java & com.github.binarywang:weixin-java-common & Merged \\
\href{https://github.com/advisories/GHSA-g5q2-cxgq-h2rw}{CVE-2020-8920} & Java & com.google.gerrit:gerrit-plugin-api & Merged \\
\href{https://github.com/advisories/GHSA-4m48-j3xj-px27}{CVE-2022-25517} & Java & com.baomidou:mybatis-plus & Non-Vuln \\
\href{https://github.com/advisories/GHSA-76x8-gg39-5jjg}{CVE-2008-0252} & Python & CherryPy & Merged \\
\href{https://github.com/advisories/GHSA-c3qv-mf8h-434r}{CVE-2008-1474} & Python & roundup & Merged \\
\href{https://github.com/advisories/GHSA-j59j-h3g7-cpmf}{CVE-2008-1475} & Python & roundup & Merged \\
\href{https://github.com/advisories/GHSA-5432-c996-hvhj}{CVE-2009-0669} & Python & ZODB3 & Merged \\
\href{https://github.com/advisories/GHSA-4849-cfqq-r8pq}{CVE-2009-2265} & Python & Products.FCKeditor & submitted \\
\href{https://github.com/advisories/GHSA-9rj9-5wcv-xgf2}{CVE-2009-2737} & Python & roundup & Merged \\
\href{https://github.com/advisories/GHSA-jqqh-999x-w26w}{CVE-2009-2959} & Python & Buildbot & Merged \\
\href{https://github.com/advisories/GHSA-mj3x-wprp-mvj9}{CVE-2009-2967} & Python & Buildbot & Merged \\
\href{https://github.com/advisories/GHSA-vvf9-jwf6-834q}{CVE-2009-3611} & Python & backintime & submitted \\
\href{https://github.com/advisories/GHSA-876c-qmcf-cxv6}{CVE-2010-0667} & Python & moin & Merged \\
\href{https://github.com/advisories/GHSA-4cm9-63x5-55wm}{CVE-2021-35958} & Python & tensorflow & submitted \\
\bottomrule
\end{tabular}
    \begin{tablenotes}
    \small
        \item ``Merged'': Its corresponding package name is accepted and merged into GitHub Advisory.
        \item ``Non-Vuln'': GitHub Advisory's maintainers do not consider it as a vulnerability.
        \item `Incorrect'': \detector{}'s output is incorrect and not accepted by maintainers.
    \end{tablenotes}
\end{threeparttable}
\vspace{-0.2cm}
\end{table*}